\documentclass[printer]{tMOP2e}

\usepackage{epsf}
\begin{document}
\doi{10.1080/09500340xxxxxxxxxxxx}
 \issn{1362-3044}
\issnp{0950-0340} %\jvol{00}
% \jnum{00} \jyear{2005} \jmonth{10 January}

\markboth{Optimal Photon Polarimeter}{Optimal Photon Polarimeter}

\title{An Optimal Photon Counting Polarimeter}

\author{Alexander Ling
%\thanks{\vspace{6pt}\newline\centerline{\tiny{ {\em phylej@nus.edu.sg}, http://www.quantumlah.org %\textcopyright 2004 Taylor \& Francis Ltd }}
%\newline\centerline{\tiny{ http://www.quantumlah.org}}
%\newline \centerline {\tiny{DOI: 10.1080/09500340xxxxxxxxxxxx}}
, Soh Kee Pang, Ant\'{\i}a Lamas-Linares, Christian Kurtsiefer}
  \received{Department of Physics, National University of Singapore, \today}
\maketitle

\begin{abstract}
We present experimental results on a method to perform polarimetry on ensembles of single photons.  Our setup is based on a measurement method known to be optimal for estimating the state of two level systems.  The setup has no moving parts and is sensitive to weak sources (emitting single photons) of light as it relies on photon counting and has potential applications in both classical polarization measurements and quantum communication scenarios.  In our implementation, we are able to reconstruct the Stokes parameters of pure polarization states with an average fidelity of 99.9\%.
\end{abstract}

\bigskip
Polarization measurement (also called polarimetry) is fundamental for many optical measurement techniques.  In classical optics, applications include stress measurement, magnetic field sensing and optical thin film characterization \cite{handbook1}.  More recently, polarization measurement of single photons prepared in particular ways has allowed experimental tests of non-locality and enabled applications in the field of quantum communication and information processing.

The purpose of this paper is to report on an implementation of an optimal polarimeter without moving parts suitable for accurate polarization measurement at faint light (single photon) levels.  Polarimetry is usually performed using a combination of rotating wave plates and a linear polarizer.  Implementations without moving parts to ensure speed and reliability have been reported before\cite{handbook2,latestfdp} and exhaustive research has been performed on optimizing them \cite{anafdp,doafdp,pol1,pol2,sabat}.  Both types of polarimeters are applicable to complete state estimation techniques (also called state tomography) which are of interest in quantum information. Optimal polarimeters have been studied recently in the context of quantum key distribution\cite{mqt,sle} where the concern is to obtain the best possible estimate on the polarization state of a limited collection of photons.  While tomographic techniques are fairly common in quantum optics and related areas \cite{qubmeas}, it should be noted that classical optics has used such techniques ever since Stokes showed how to determine completely the polarization of an unknown state of light \cite{refstokes}.

One instrumentally motivated way to fully characterize the polarization state of light uses three (Stokes) parameters $S_x, S_y$ and $S_z$.  The three parameters may be augmented by an intensity $S_m$ and together form a Stokes vector, which can be normalized to be $\vec{S} = (1,S_x/S_m,S_y/S_m,S_z/S_m)$; a reduced Stokes vector $\vec{S_r} = (S_x,S_y,S_z)/S_m$ is used to identify a point in the Poincare sphere.  

%Direct Stokes polarimeters take six detector readings to determine three parameters where each parameter is determined by two detector readings.  The projection of the input polarization state on the conjugate vectors making up the three axes of the Poincare sphere determine these detector outcomes.  Corresponding detector outcomes are subtracted to obtain components of the Stokes vector.  The conjugate vectors are horizontal and vertical linear polarization for the $x-$axis, $\pm45^{\circ}$ linear polarization for the $y-$axis and circular polarization for the $z-$axis.    

A minimal scheme of measurement requires only four detector readings to estimate the Stokes parameters.  This corresponds to determining the overlap of the Stokes vector with four non-coplanar vectors in the Poincare Sphere.  It has been shown that if this vector quartet defines a tetrahedron, then it has the necessary conditions for optimal polarimetry \cite{anafdp,pol2,sabat}.  Independently, \u{R}eh\'{a}\u{c}ek et al showed that this is also true for two level quantum systems \cite{mqt}.   Optimal here is defined as having the least uncertainty for determining polarization from a given amount of light energy and that all possible input states are determined equally well.  In the context of photon counting, it means that the maximum amount of polarization information is extracted per photon counted.  Thus, polarimeters based on a tetrahedron measurement are optimal and not only minimal. 

Independent of previous descriptions of four detector polarimeters, we designed and implemented a division-of-amplitude polarimeter based on the tetrahedron measurement.  The setup is shown in Fig 1.  This works by dividing the incoming light beam using a partially polarizing beam splitter (PPBS).  Polarizing beam splitters are then used to analyze the daughter beams in different polarization bases.  The design goal was to optimize the beam splitting ratio of the PPBS; optimizing this gives us an optimal polarimeter.

Although previous reports describe division-of-amplitude polarimeters with fewer optical components \cite{handbook2} (no quartz compensators and no waveplates), they are often less than optimal tetrahedron measurements.  Furthermore, these original designs place stringent demands on the performance of the PPBS.  It must rotate light leaving the beamsplitter into the correct polarization basis as well as preserve the phase relationship between the horizontal (H) and vertical (V) polarizations.  

In our experiment we relaxed the requirement of having minimal optical components and aimed only to achieve optimal tetrahedron measurements without moving parts.  We used a non-ideal PPBS, specifying only an intensity splitting ratio required to work with selected waveplates (an ideal PPBS would not need waveplates and compensators but discussion of how to build such a device is out of the scope of this report; work continues in this direction \cite{idppbs}).  Quartz compensators were used to correct for any relative phase shifts between the H and V polarizations introduced by the PPBS.  Although a non-ideal PPBS requires us to have extra components they do not increase the complexity of the setup by much; in fact it is somewhat simpler because we are able to work with incident and exit angles of $90^\circ$ for most of our elements.  

To understand how we found the optimal beamsplitting ratio, let us associate each detector with a vector $\vec{b_j}$ from the tetrahedron quartet.  Each vector $\vec{b_j}$ is related to a measurement outcome $I_j$ by \begin{eqnarray} I_j   =  \frac{I_t}{4} (\vec{b_j} \cdot \vec{S}), \quad I_t = \sum_j I_j\end{eqnarray} where $I_j$ is the intensity of light falling on the associated detector $j$ (for single photon counters, $I_j$ refers to the photon count rate).  The intensities may be expressed as a vector $\vec{I} = (I_1, I_2, I_3, I_4)$, so we have $\vec{I} = B \cdot \vec{S}$ with an instrument matrix $B$ that relates a Stokes vector, $\vec{S}$ to measured outcomes.  When light of polarization $\vec{b_j}$ is used, the normalized intensity of light falling on detector $j$ is $\frac{1}{2}$.   At all other detectors, the intensity is  $\frac{1}{6}$.  Similarly, when light with polarization conjugate to $\vec{b_j}$ is introduced, no light arrives at detector $j$ but is split between all other detectors equally.   
%To understand the optimization scheme, we associate each detector with a vector $\vec{b_j}$ from the tetrahedron quartet.  Each vector $\vec{b_j}$ is related to a measurement outcome $I_j$ by \begin{eqnarray} I_j   =  \frac{I_t}{4} (\vec{b_j} \cdot \vec{S}), \quad I_t = \sum_j I_j\end{eqnarray} where $I_j$ is the intensity of light falling on the associated detector $j$ (if we use single photon counters, $I_j$ refers to the photon count rate).  The intensities may be expressed as a vector $\vec{I} = (I_1, I_2, I_3, I_4)$, so we have $\vec{I} = B \cdot \vec{S}$, and we obtain an instrument matrix $B$ that relates a Stokes vector to measured outcomes.  When light of polarization $\vec{b_j}$ is used, photons arrive at detector $j$ with a relative frequency of $\frac{1}{2}$ and at all other detectors, the relative frequency is  $\frac{1}{6}$.  Similarly, when light polarized orthogonal to $\vec{b_j}$ is introduced, no light arrives at detector $j$ but is split between all other detectors equally.   

\begin{figure}
\centerline{{\epsfxsize=5cm \epsfbox{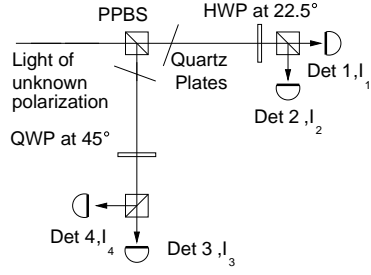}}} \caption{Setup for optimal polarimetry. The partially polarizing beam splitter (PPBS) light splitting ratio is important for our selection of waveplate angles.  The quartz plates remove the relative phase difference between the horizontal and vertical polarizations introduced by the PPBS.  In both transmitted and reflected arms, polarizing beam splitters then divide the light further between two detectors.  Note that beams enter and leave the beamsplitter faces at $90^\circ$ so alignment is straightforward.} \label{fig2} 
\end{figure}

The optimal splitting ratio for the PPBS is then most easily determined using Jones vectors (or the equivalent spin-$\frac{1}{2}$ vectors for polarization qubits).  Following Figure 2, suppose the PPBS has the property that input light with the polarization state $ \alpha \choose \beta$  leads to the polarizations $y\alpha \choose x\beta$ and $x\alpha \choose y\beta$ in the transmitted and reflected arms, respectively.  Two orthogonal vectors may be expressed as $\cos\theta \choose e^{i\phi} \sin\theta$ and $-e^{-i\phi}\sin\theta \choose \cos\theta$.  Thus, the normalized intensity of light falling on detector $j$ (e.g. $j=1,2)$ is \begin{eqnarray} I_1/I_t &=& \left| \left( \begin{array}{cc} \cos\theta & e^{-i\phi}\sin\theta \end{array}\right) {\alpha y \choose \beta x}\right|^2\nonumber\\ &=&\left|\alpha y \cos\theta + \beta x e^{-i\phi} \sin\theta\right|^2 \\ I_2/I_t &=& \left|-\alpha y e^{i \phi} \sin\theta + \beta x \cos\theta\right|^2 \end{eqnarray} Choosing a different measurement basis for detectors 3 and 4, we get: \begin{eqnarray} I_3/I_t &=& \left|\alpha x \cos\theta ' + \beta y e^{-i \phi '} \sin\theta '\right|^2 \\ I_4/I_t &=& \left|-\alpha x e^{i \phi '} \sin\theta ' + \beta y \cos\theta '\right|^2.  \end{eqnarray}
%The optimal splitting ratio for the PPBS is then most easily determined using Jones vectors (or the equivalent spin-$\frac{1}{2}$ vectors for polarization qubits).  Following Figure 2, suppose the PPBS has the property that input light with the polarization state $ \alpha \choose \beta$  leads to the polarizations $y\alpha \choose x\beta$ and $x\alpha \choose y\beta$ in the transmitted and reflected arms respectively.  Two orthogonal vectors may be expressed as $\cos(\theta) \choose e^{i\phi} \sin(\theta)$ and $-e^{-i\phi}\sin(\theta) \choose \cos(\theta)$.  Thus, the probability of an input photon reaching any detector $j$ (e.g. $j=1,2)$ is \begin{eqnarray} \mbox{prob}(j = 1) &=& \left| \left( \begin{array}{cc} \cos(\theta) & e^{-i\phi}\sin(\theta) \end{array}\right) {\alpha y \choose \beta x}\right|^2\\ &=&\left|\alpha y \cos(\theta) + \beta x e^{-i\phi} \sin(\theta)\right|^2 \\ \mbox{prob}(j = 2) &=& \left|-\alpha y e^{i \phi} \sin(\theta) + \beta x \cos(\theta)\right|^2 \end{eqnarray} Choosing a different measurement basis for detectors 3 and 4, we get: \begin{eqnarray} \mbox{prob}(j=3) &=& \left|\alpha x \cos(\theta ') + \beta y e^{-i \phi '} \sin(\theta ')\right|^2 \\ \mbox{prob}(j=4) &=& \left|-\alpha x e^{i \phi '} \sin(\theta ') + \beta y \cos(\theta ')\right|^2.  \end{eqnarray}

Since the tetrahedron can be oriented arbitrarily, we choose for convenience $45^{\circ}$ linear polarization ($\theta=\pi/4,\quad \phi=0$) in the transmitted arm, and circular polarization ($\theta '= \pi/4 \quad \phi '=\pi/2$) in the reflected arm.  Using these values we find splitting ratios of \begin{eqnarray} x^2 = \frac{1}{2} + \frac{1}{2\sqrt{3}} ,\quad y^2= \frac{1}{2} - \frac{1}{2\sqrt{3}}\end{eqnarray}  The same result was shown independently by maximizing the instrument matrix determinant\cite{doafdp}.  Beam splitters with this nominal splitting ratio were obtained from a commercial optical coating facility. 

To use this polarimeter, the instrument matrix is determined by calibration.  For this purpose, we prepared input light with four different polarizations whose Stokes vector are non-coplanar, giving a set of four simultaneous equations which are used to determine the elements in the matrix, $B$.  This matrix takes into account losses and any non-ideal splitting ratio $x,y$ of the PPBS.  As a choice of calibration vectors, we select vectors that are maximally far apart in the Poincare sphere.  The natural choice for this is again a vector quartet defining a tetrahedron.  The normalized Stokes vector representation for our selected vector quartet used to calibrate our polarimeters are 
\begin{eqnarray}
\left. \begin{array}{c} \vec{b_1} \\ \vec{b_2} \end{array}
\right\}  =  \left( \begin{array}{c}
  1 \\ \sqrt{\frac{1}{3}}\\ \pm \sqrt{\frac{2}{3}} \\0
\end{array}
\right), \quad
\left. \begin{array}{c} \vec{b_3} \\ \vec{b_4} \end{array}
\right\}  =  \left( \begin{array}{c}
1 \\  -\sqrt{\frac{1}{3}} \\ 0 \\ \mp \sqrt{\frac{2}{3}}
\end{array}
\right)
\end{eqnarray}
%\subsection{}
%\subsubsection{}

In anticipation of future experiments, we prepared these states using light at 702 nm from a spontaneous parametric down conversion source \cite{kwiat95}.   The down converted light was collected into single mode optical fibres and has a spectral bandwidth of 4.74 $\pm$ 0.04 nm \cite{higheff}.  We prepared the tetrahedron vector quartet using a Glan-Thomson polarizer (extinction ratio of $10^5$) followed by a halfwave plate (HWP) and a quarterwave plate (QWP) allowing an arbitrary pure polarization state to be generated for calibration.  Light detection is performed by passively quenched Silicon avalanche photodiodes. 

By this method the instrument matrix inverse was found to be
\begin{eqnarray}
B^{-1} = \left(
\begin{array}{cccc}
1.015 & 0.988 & 0.981 & 1.013 \\
2.779 &-3.156 & 0.231 & 0.302 \\
-0.138 & 0.149 & -3.309 & 3.493\\
2.004 & 1.458 & -1.784 & -1.865
\end{array}
\right)
\end{eqnarray}

The uncertainty in the elements has a maximum value of $\pm$ 0.007 deduced from propagated Poissonian counting statistics.  Here we are limited only by the final accuracy of the state preparation apparatus.  

With the instrument matrix determined, it is possible to reconstruct the Stokes vector of any input state.  This was tested by preparing a large number of pure states using our polarization state generator.  For each input state, the Stokes vector was reconstructed from the detected outputs.  We characterize the accuracy of the measurement by a fidelity $F$ of the reconstructed state to the prepared state.  The fidelity for two states is defined as $F=\{\mbox{Tr}[\sqrt{\sqrt{\sigma_{th}}\rho_{rec}\sqrt{\sigma_{th}}}]\}^2$, where $\sigma_{th} \mbox{ and } \rho_{rec}$ are the coherency matrices (or density matrices) of the theoretical and reconstructed states, respectively \cite{fidelity}.  For pure states, this reduces to $\frac{1}{2}(\vec{S_{rec}}\cdot \vec{S_{th}})$, i.e. the overlap of their normalized Stokes vectors.  

The map of this fidelity is shown in Figure 2.  The average fidelity is 99.9\% with a smallest value of 99.4 $\pm$ 0.4 \% for circularly polarized light.  This is attributed to imperfections in the waveplates of our state generator.  It should be noted, however, that the range in fidelity can be accounted for by approximately 4 mrad alignment error in the waveplates which is comparable with the specified accuracy of 5 mrad for the control motors. 

A careful inspection of Figure 2 shows that this optimal polarimeter has no large areas of low fidelity showing that it estimates all input states equally well.  This is of particular importance in low light applications with single photon sources.  In our results, cumulative photon counts at each measured point was on the order of $10^5$ (approximately 0.028 pJ).  
%While this exactly a low number of photons, the fidelity obtained at each point is already at the asymptotic limit estimated by J.~\u{R}eh\'{a}\u{c}ek et al \cite{mqt}.  
Further work to determine the asymptotic efficiency of tetrahedron polarimeters is currently planned.

\begin{figure}
\centerline{\epsfxsize=6cm \epsfbox{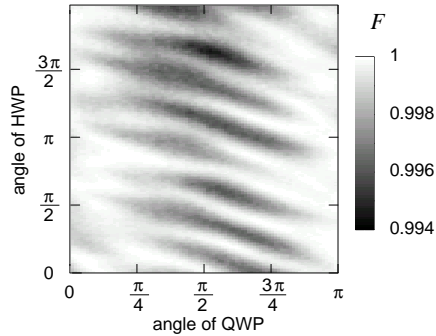}} \caption{Fidelity($F$) of reconstructed Stokes Vectors.  The average fidelity is $99.9\%$ while the minimum fidelity is $99.4\%$.  Input states were generated from H polarized light transmitted through a half-wave plate and a quarter-wave plate.} 
\end{figure}

In conclusion, we have presented an experimental realisation of optimal polarimetry by photon counting. In the context of quantum information, this is a realisation of single qubit tomography based on the tetrahedron measurement (that is known to be optimal \cite{mqt}).  A pair of these polarimeters together could be used to perform state tomography of entangled photon pairs.

The setup is very stable and fast as it has no moving parts.  This could be applied to scenarios requiring time-resolved polarimetry or advanced quantum communication schemes, where optimal measurements on qubits can enhance communication bandwidth \cite{sprot}.  In its current form, the setup is able to reconstruct the Stokes vector of any input polarization to at least a fidelity of 99.4\%.  Although the work is presented for pure polarization states, there is no reason to believe it will not work for partially polarized light (called mixed polarization states) to be addressed in future work.   

We would like to thank B.-G. Englert and Janet Anders for helpful discussions.  This work was supported by DSTA Grant No. R-394-000-019-422 and A*Star Grant No. R-144-000-071-305.

\end{document}